\documentclass[article]{elsarticle}
\usepackage{a4wide}
\usepackage{natbib} 			
\usepackage{graphicx}
\usepackage{amsfonts}
\usepackage{amsmath}
\usepackage{booktabs}
\usepackage{epsfig}
\usepackage{hyperref}
\hypersetup{citecolor=black,linkcolor= black}

\begin{document}
\title{Finite sample properties of power-law cross-correlations estimators}
\author{Ladislav Kristoufek}
\ead{kristouf@utia.cas.cz}
\address{Institute of Information Theory and Automation, Academy of Sciences of the Czech Republic, Pod Vodarenskou Vezi 4, 182 08, Prague 8, Czech Republic\\
Institute of Economic Studies, Faculty of Social Sciences, Charles University, Opletalova 26, 110 00, Prague 1, Czech Republic
}

\begin{abstract}
We study finite sample properties of estimators of power-law cross-correlations -- detrended cross-correlation analysis (DCCA), height cross-correlation analysis (HXA) and detrending moving-average cross-correlation analysis (DMCA) -- with a special focus on short-term memory bias as well as power-law coherency. Presented broad Monte Carlo simulation study focuses on different time series lengths, specific methods' parameter setting, and memory strength. We find that each method is best suited for different time series dynamics so that there is no clear winner between the three. The method selection should be then made based on observed dynamic properties of the analyzed series.
\end{abstract}


\journal{Physica A}

\maketitle


\section{Introduction}

Power-law cross-correlations have become a popular and frequently analyzed topic in various disciplines covering seismology \cite{Shadkhoo2009}, hydrology \cite{Hajian2010}, (hydro)meteorology \cite{Vassoler2012,Kang2013}, biology \cite{Xue2012},  biometrics \cite{Ursilean2009}, DNA sequences \cite{Stan2013}, neuroscience \cite{Jun2012}, electricity \cite{Wang2013b}, finance \cite{Podobnik2009a,Lin2012,Shi2013}, commodities \cite{He2011,He2011a}, traffic \cite{Zebende2009,Xu2010,Zhao2011}, geophysics \cite{Marinho2013} and others. The analysis is standardly based on an estimation of the bivariate Hurst exponent $H_{xy}$ which is connected to an asymptotic power-law decay of the cross-correlation function or a divergent (again following a power-law) at origin cross-power spectrum. Specifically, a power-law cross-correlated process has the cross-correlation function of a form $\rho_{xy}(k) \propto k^{2H_{xy}-2}$ for lag $k \rightarrow +\infty$ and the cross-power spectrum of a form $|f_{xy}(\lambda)|\propto \lambda^{1-2H_{xy}}$ for frequency $\lambda \rightarrow 0+$. In a similar way as for the univariate case, the bivariate Hurst exponent of 0.5 is characteristic for no power-law cross-correlations. Processes with $H_{xy}>0.5$ are then cross-persistent and they tend to move together whereas for $H_{xy}<0.5$ they are more likely to move in opposite directions.

Most of the literature focusing on power-law cross-correlations is empirical and there are no studies of statistical properties of the utilized estimators. Here, we try to fill this gap and we present a broad Monte Carlo simulation study of performance of three popular bivariate Hurst exponent estimators -- detrended cross-correlation analysis \cite{Podobnik2008,Zhou2008,Jiang2011}, height cross-correlation analysis \cite{Kristoufek2011} and detrending moving-average cross-correlation analysis \cite{Arianos2009,He2011a}. Specifically, we focus on an ability of the estimators to precisely estimate the bivariate Hurst exponent not only under a simple setting of standard power-law cross-correlations when the bivariate Hurst exponent exponent equals to an average of the Hurst exponent of the separate processes but also under potential short-term memory bias and under power-law coherency. The paper is organized as follows. In Section 2, we introduce all three analyzed estimators. In Section 3, the Monte Carlo simulation setting is described. In Section 4, the results are presented in detail. Section 5 concludes.

\section{Methodology}

\subsection{Detrended cross-correlation analysis}

Detrended cross-correlation analysis (DCCA, or DXA) is the most frequently used method for the estimation of the bivariate Hurst exponent in the time domain. Podobnik \& Stanley \cite{Podobnik2008} construct the method as a bivariate generalization of the detrended fluctuation analysis (DFA), which is again probably the most popular heuristic method of estimating the (generalized) Hurst exponent \citep{Peng1993,Peng1994,Kantelhardt2002}. DCCA was further generalized for the multifractal analysis by Zhou \cite{Zhou2008} and the multifractal detrended cross-correlation analysis (MF-DXA) was developed. Jiang \& Zhou \cite{Jiang2011} altered the filtering procedure in MF-DXA with using the moving averages to create the multifractal detrending moving average cross-correlation analysis (MF-X-DMA). DCCA was also used to construct statistical tests for the presence of long-range cross-correlations between two series \citep{Zebende2011,Podobnik2011,Balocchi2013,Blythe2013,Zebende2013,Kristoufek2014}.

In the DCCA procedure, we consider two long-range cross-correlated series $\{x_t\}$ and $\{y_t\}$ with $t=1,\ldots,T$. Their respective profiles $\{X_t\}$ and $\{Y_t\}$, defined as $X_t=\sum_{i=1}^t{x_i-\bar{x}}$ and $Y_t=\sum_{i=1}^t{y_i-\bar{y}}$, for $t=1,\ldots,T$, are divided into overlapping boxes of length $s$ so that $T-s+1$ boxes are constructed. In each box between $j$ and $j+s-1$, the linear fit of a time trend is constructed so that we get $\widehat{X_{k,j}}$ and $\widehat{Y_{k,j}}$ for $j\le k \le j+s-1$. The covariance between residuals in each box is defined as
\begin{equation}
f_{DCCA}^2(s,j)=\frac{\sum_{k=j}^{j+s-1}{(X_k-\widehat{X_{k,j}})(Y_k-\widehat{Y_{k,j}})}}{s-1}.
\label{eq:DCCA1}
\end{equation}
The covariances are finally averaged over the blocks of the same lengths $s$ and the detrended covariance is obtained as
\begin{equation}
F_{DCCA}^2(s)=\frac{\sum_{j=1}^{T-s+1}{f_{DCCA}^2(s,j)}}{T-s}.
\label{eq:DCCA2}
\end{equation}
For the long-range cross-correlated processes, the covariance scales as
\begin{equation}
F_{DCCA}^2(s)\propto s^{2H_{xy}}.
\label{eq:DCCA3}
\end{equation}

The estimate of the bivariate Hurst exponent is obtained by the log-log regression on Eq. \ref{eq:DCCA3}. Similarly to DFA and MF-DFA, there are several ways of treating overlapping and non-overlapping boxes of length $s$, compare e.g. Refs. \cite{Peng1993,Kantelhardt2002,Taqqu1995,Barunik2010,Kristoufek2010,Grech2013,Grech2013a}. In the simulations, we use non-overlapping boxes with a step between $s$ equal to 10 due to computational efficiency.

\subsection{Height cross-correlation analysis}

Kristoufek \cite{Kristoufek2011} introduces the multifractal height cross-correlation analysis (MF-HXA) as a bivariate generalization of the height-height correlation analysis \citep{Barabasi1991a,Barabasi1991b,Alvarez-Ramirez2002} and the generalized Hurst exponent approach \citep{DiMatteo2003,DiMatteo2005,DiMatteo2007}, which are often labeled simply as HHCA and GHE, respectively. 

MF-HXA is constructed to analyze the multifractal properties of bivariate series similarly to MF-DXA. We generalize the $q$-th order height-height correlation function for two simultaneously recorded series. Let us consider two profiles $\{X_t\}$ and $\{Y_t\}$ with time resolution $\nu$ and $t=\nu,2\nu,...,\nu\lfloor\frac{T}{\nu}\rfloor$, where $\lfloor \rfloor$ is a lower integer sign. For better legibility, we denote $T^{\ast}=\nu\lfloor\frac{T}{\nu}\rfloor$, which varies with $\nu$, and we write the $\tau$-lag difference as $\Delta_{\tau}X_t \equiv X_{t+\tau}-X_t$ and $\Delta_{\tau}X_tY_t \equiv \Delta_{\tau}X_t\Delta_{\tau}Y_t$. For analysis of power-law cross-correlations, i.e. the case when $q=2$, the height-height covariance function is then defined as

\begin{equation}
\label{eq:MFHXAeq1}
K_{xy,2}(\tau)=\frac{\nu}{T^{\ast}}\sum_{t=1}^{T^{\ast}/\nu}|\Delta_{\tau}X_tY_t| \equiv \langle|\Delta_{\tau}X_tY_t|\rangle
\end{equation}
where time interval $\tau$ generally ranges between $\nu=\tau_{min},\ldots,\tau_{max}$. Scaling relationship between $K_{xy,q}(\tau)$ and the generalized bivariate Hurst exponent $H_{xy}(q)$ is obtained as

\begin{equation}
\label{eq:MFHXAeq2}
K_{xy,2}(\tau) \propto \tau^{H_{xy}}.
\end{equation}

Obviously, MF-HXA reduces to the height-height correlation analysis of Barabasi \textit{et al.} \cite{Barabasi1991a,Barabasi1991b} for $\{X_t\}=\{Y_t\}$ for all $t=1,\ldots,T$. Note that it makes sense to analyze the scaling according to Eq. \ref{eq:MFHXAeq2} only for detrended series $\{X_t\}$ and $\{Y_t\}$ \citep{DiMatteo2007}. A type of detrending can generally take various forms --  polynomial, moving averages and other filtering methods -- and is applied for each time resolution $\nu$ separately. 

The estimated bivariate Hurst exponent is again obtained via the log-log regression. It has been argued that the best estimates and the most regular scaling is obtained for $\tau/T\rightarrow 0$ \citep{Barabasi1991a,Barabasi1991b} and it has been shown that the most appropriate setting is to use a fixed $\tau_{min}=1$ and several values of $\tau_{max}$, usually between 5 and 19 (or 20), and take the average Hurst exponent of these estimates as the best fit to the actual value, which practically means obtaining the jackknife estimate of Hurst exponent \cite{DiMatteo2003,DiMatteo2005,DiMatteo2007,Barunik2010,Kristoufek2010a,Barunik2012}.

\subsection{Detrended moving-average cross-correlation analysis}

Detrending moving average (DMA) was proposed as a method for estimating Hurst exponent by Alessio \textit{et al.} \cite{Alessio2002} motivated by the work of Vandewalle \& Ausloos \cite{Vandewalle1998}. Even though the method is not directly connected to the power-law decay of auto-correlations nor to the scaling of variances of the partial sums nor the diverging power spectrum, it has been frequently applied mainly due to its computational efficiency. The connection between the estimator itself and the actual long-range dependence -- that the variance of integrated series of the long-range dependent process follows a power-law with respect to the length of the moving window -- has been shown numerically \citep{Grech2005,Barunik2010,Grech2013b}.

To check whether the relationship holds also for the scaling of covariances, He \& Chen \cite{He2011a} proposed a new method called detrended moving-average cross-correlation analysis (DMCA) as a special case of the method of Arianos \& Carbone \cite{Arianos2009}. Note that this should not be confused with the MF-X-DMA method of Jiang \& Zhou \cite{Jiang2011}, which applies the moving average filtering to the DCCA or MF-DXA methodology, or with 2D-DMA of Carbone \cite{Carbone2007}.

For two series $\{x_t\}$ and $\{y_t\}$ and their respective profiles $\{X_t\}$ and $\{Y_t\}$, the detrended covariance $F_{DMCA}^2(\kappa)$ is defined as
\begin{equation}
\label{eq:CC-DMA}
F_{DMCA}^2(\kappa)=\frac{1}{T-\kappa+1}\sum_{i=\lfloor\kappa/2\rfloor+1}^{T-\lfloor\kappa/2\rfloor}{\Big(X_i-\widetilde{X_i(\kappa)}\Big)\Big(Y_i-\widetilde{Y_i(\kappa)}\Big)},
\end{equation}
where $\widetilde{X_i(\kappa)}$ and $\widetilde{Y_i(\kappa)}$ are respective non-weighted centered moving averages at time point $i$ with a moving average window of length $\kappa=1,3,5,\ldots,\kappa_{max}$, where $\kappa_{max}$ is an odd integer. In general, the specification of the moving average can take various forms (centered, backward, forward, weighted or unweighted). For the long-range cross-correlated processes $\{x_t\}$ and $\{y_t\}$, we expect to observe
\begin{equation}
\label{eq:CC-DMA_scaling}
F_{DMCA}^2(\kappa)\propto \kappa^{2H_{xy}}.
\end{equation}
Note that compared to DCCA, the series is not split into boxes which makes the method much more straightforward and computationally efficient.

\section{Monte Carlo simulation setting}

We are primarily interested in three aspects of the bivariate Hurst exponent estimation. Firstly, we examine a standard power-law cross-correlations case when the bivariate Hurst exponent equals the average of the separate Hurst exponents. Secondly, we focus on a potential bias caused by the short-term memory. And thirdly, we study the case when the bivariate Hurst exponent is lower than the average of the separate exponents, i.e. the power-law coherency case (see Sela \& Hurvich \cite{Sela2012} for more details).

For the standard power-law cross-correlations case, we utilize correlated ARFIMA processes. Specifically, the processes $\{x_t\}$ and $\{y_t\}$ are defined as
\begin{gather}
\label{eq:ARFIMA1}
x_t=\sum_{n=0}^{\infty}{a_n(d_1)\varepsilon_{t-n}} \nonumber\\
y_t=\sum_{n=0}^{\infty}{a_n(d_2)\nu_{t-n}}
\end{gather}
where
\begin{gather}
\label{eq:a}
a_n(d)=\frac{\Gamma(n+d)}{\Gamma(n+1)\Gamma(d)}.
\end{gather}
Specifically, we use the ARFIMA processes with $d_1=d_2=d$ with correlated error terms of level $\rho_{\varepsilon\nu}$\footnote{For a given correlation level $\rho$, the correlated series $X$ and $Y$ are constructed using surrogate series $Z$ in the following way. Series $X$ and $Z$ are generated as independent standard Gaussian processes. The series $Y$ is then formed as $Y=\rho X + \sqrt{1-\rho^2}Z$. This way, we obtain standard Gaussian series $X$ and $Y$ with a pairwise correlation of $\rho$.}. The theoretical bivariate Hurst exponent is then equal to $H_{xy}=d+0.5$. In the simulation study, we analyze two cases -- $H_x=H_y=0.6$ and $H_x=H_y=0.9$ -- to cover both weak and strong power-law cross-correlations. 

For the short-term memory bias examination, we use a combination of a long-term dependent ARFIMA process and a short-term dependent AR(1) process so that we have
\begin{gather}
x_t=\sum_{n=0}^{\infty}{a_n(d_1)\varepsilon_{t-n}} \nonumber \\
y_t=\theta y_{t-1}+\nu_t=\sum_{n=0}^{\infty}{\theta^n\nu_{t-n}}
\label{eq:ARFIMAAR}
\end{gather}
with $|\theta|<1$. We again work with correlated error terms. In this case, the theoretical bivariate Hurst exponent is equal to $H_{xy}=\frac{1}{2}d_1+0.5$ as the process $\{y_t\}$ is only short-term correlated \cite{Kristoufek2014arXiv}. In the Monte Carlo study, we fix $H_x=0.9$ and vary $\theta=0.1,0.5,0.8$ to control for weak, medium and strong short-term memory.

For the last case, we utilize the mixed correlated ARFIMA processes \cite{Kristoufek2013} defined as
\begin{gather}
\label{eq:ARFIMA_LC}
x_t=\alpha\sum_{n=0}^{\infty}{a_n(d_1)\varepsilon_{1,t-n}}+\beta\sum_{n=0}^{\infty}{a_n(d_2)\varepsilon_{2,t-n}} \nonumber \\
y_t=\gamma\sum_{n=0}^{\infty}{a_n(d_3)\varepsilon_{3,t-n}}+\delta\sum_{n=0}^{\infty}{a_n(d_4)\varepsilon_{4,t-n}}.
\end{gather}
where only the error terms $\{\varepsilon_2\}$ and $\{\varepsilon_3\}$ are correlated, and the other pairs are uncorrelated. If we set $d_1>d_2$ and $d_4>d_3$, we obtain the power-law coherency case when $H_{xy}<(H_x+H_y)/2$. Specifically, we have $H_x=d_1+0.5$, $H_{y}=d_4+0.5$ and $H_{xy}=0.5+\frac{1}{2}(d_2+d_3)$. In the simulation study, we set $d_1=d_4=0.4$ and $d_2=d_3=0.2$ so that the theoretical Hurst exponents are equal to $H_x=H_y=0.9$ and $H_{xy}=0.7$.

In all cases, we analyze three different time series lengths -- $T=500,1000,5000$ -- and we are interested in the effect of the strength of the correlation between the error terms. To do so, we check the finite sample properties for the correlation levels 0.1, 0.5 and 0.9. Additionally, we examine two variants of each estimator -- both with and without absolute values in there respective covariance definitions as these slightly vary across the literature. Other parameters of the simulations are specified for each estimator later if necessary. Overall, we are interested in their bias, standard error and mean squared error.

\section{Results}

\subsection{Detrended cross-correlation analysis}
Statistical performance of detrended fluctuation analysis has been shown to be dependent on a choice of $s_{min}$ and $s_{max}$, i.e. the minimum and the maximum scales taken into consideration \citep{Weron2002,Grech2005,Kristoufek2010,Barunik2010}. In the same way, DCCA can perform differently for various settings. We set $s_{max}=T/5$, which is standardly done in the literature and we manipulate $s_{min}=10,20,50$ for $T=500$ and $s_{min}=10,50,100$ for the other two cases, $T=1000,5000$. The results of the simulations for DCCA are summarized in \autoref{tab:DCCA1}-\ref{tab:DCCA6_abs}.

For ARFIMA processes with correlated error terms, there are several interesting findings (\autoref{tab:DCCA1}-\ref{tab:DCCA2}). First, both the original and the absolute value based definitions of DCCA are biased downwards, while the absolute value based version is less biased than the original one. Second, standard deviation of the estimators increases with the increasing $s_{min}$. Third, for the short series of $T=500$, the standard deviations of the original DCCA are much higher than for the absolute value version. Fourth, the original version of DCCA is practically useless for weakly correlated processes as for the case when the correlation between error terms is only 0.1, there are so many cases when the log-log regression could not be performed that it does not make sense to report the results. We shall see that this is true for all the studied methods. Fifth, for the absolute value version of the estimator, the standard deviation increases with the correlation between error terms, which is probably caused by the fact that the absolute value version of the estimator is pushed towards the average of the separate Hurst exponents and the correlation between error terms only disturbs this effect. Sixth, the inverse is true for the original version of DCCA, i.e. the variance of the estimator decreases with the correlation between error terms. Seventh, bias and variance of the estimator decreases with time series length. And eighth, mean squared error decreases with the time series length.

For long-range cross-correlations arising from the combination of ARFIMA and AR(1) processes, the results are summarized in \autoref{tab:DCCA3}-\ref{tab:DCCA5}. In general, both DCCA-based estimators are quite robust to the presence of short-range dependence even for a very strong memory of $\theta=0.8$. More specific findings follow. First, for the weak short-term memory ($\theta=0.1$), both estimators are still biased downwards, which is more profound for the original DCCA procedure. Interestingly, the mean squared error decreases with increasing $s_{min}$ for both estimators which is rather unexpected as the short-range dependence usually mostly affects the lower scales. Variance of the original estimator is higher than for the absolute value version of the estimator. Again, the original DCCA estimator performs very poorly for the lowest values of correlation between error terms and it is not reported for any of the processes. Second, for the processes with more profound short memory, we observe an expected behavior of bias -- it reduces with the increasing minimum scale $s_{min}$. However, the variance of the estimator increases with $s_{min}$ so that it more than offsets the bias gains and the total mean squared error increases with the increasing $s_{min}$ for both versions of the estimator. In terms of a choice of the estimator and parameter setting, we would suggest to use low minimum scale for both versions of the estimator as it provides a good balance between bias and variance. Note that for $T=5000$, $\theta=0.8$ and the strongest correlation between error terms, the mean squared errors of the estimator equal to 0.0033 and 0.0036 for $s_{min}=10$ for the original version and the absolute value version of the estimator, respectively, and does not increase markedly for the other minimum scales, and the DCCA estimators are thus very robust to the potential short-term memory bias.

The situation changes for the mixed-correlated ARFIMA processes (\autoref{tab:DCCA6_abs}). Disturbingly, the estimates based on the original version of DCCA are not able to estimate $H_{xy}$ very frequently even for the correlation between error terms of 0.5, due to the number of cases when the detrended covariances are negative for several scales. We thus provide the results only for the absolute value version of DCCA. First, the bias increases with the time series length and the estimator is thus not consistent. Second, the bias and variance, and thus also mean squared error, of the estimator increase with increasing $s_{min}$. Third, the bias decreases with the strength of correlations between error terms. Unfortunately, this decrease is only mild. The estimator is thus not a very good tool to estimate $H_{xy}$ if it is not equal to the average of the separate Hurst exponents.

\subsection{Height cross-correlation analysis}

For height cross-correlation analysis, we are mainly interested in its performance using various maximum scales $\tau_{max}$. As a starting point, we use $\tau^{\ast}_{max}=20$, which is frequently suggested in the literature \citep{DiMatteo2003,DiMatteo2005,DiMatteo2007,Kristoufek2010a,Kristoufek2011}. Moreover, we check the maximum scales of 50 and 100. To obtain more stable estimates of the bivariate Hurst exponent, we apply the jackknife procedure which estimates the Hurst exponent as a mean of Hurst exponents estimated using $\tau_{min}=1$ and $\tau_{max}=5,\ldots,\tau^{\ast}_{max}$. The simulated processes are the same as for DCCA methods and the results are summarized in \autoref{tab:HXA1}-\ref{tab:HXA6}.

For correlated ARFIMA processes (\autoref{tab:HXA1}-\ref{tab:HXA2}), we observe that both versions of the estimator are downward biased and the bias is more severe for stronger long-range cross-correlations. For lower levels of correlations, i.e. the lower correlations of error terms, the original (absolute value based) approach strongly outperforms the adjusted version. However, the differences become negligible for the error terms correlation of 0.9. In general, it holds that the bias is more severe and the variance of the estimators increase with increasing $\tau^{\ast}_{max}$. Similarly to the DCCA estimators, we find that the behavior of variance of the estimators differs for the two approaches. For the absolute value based HXA, the variance slightly increases with the correlation of error terms while the opposite is true for the alternative specification, which is indeed more desirable and intuitive. The interpretation is the same as for DCCA -- the absolute value approach to the time domain approaches pushes the estimates of the bivariate Hurst exponent to the average of the separate Hurst exponents, which is obviously not desirable. Nonetheless, the bias and variance of both versions of the estimator decrease with the time series length.

The HXA method is quite robust to the short memory bias as is shown for the combination of ARFIMA and AR(1) processes in \autoref{tab:HXA3}-\ref{tab:HXA5}. The properties of the estimators depend on the strength of the short memory component. For weak short memory ($\theta=0.1$), the properties are practically the same as for the previous cases -- bias and variance increase with $\tau_{max}$ but decrease with the time series length with the same differences between the original and alternative definitions of the method. For the stronger short memory, we observe that even though the variance of the estimator still increases with $\tau_{max}$, the bias decreases. Such disproportion is the most obvious for the longest analyzed time series length for which the mean squared error decreases with the increasing $\tau_{max}$. Interestingly, the adjusted HXA outperforms the original HXA in a sense of the mean squared error for the longest series length and the strongest correlation between error terms. For the strong short memory with $\theta=0.8$, both estimators are strongly upwards biased and the bias increases with the time series length for both specifications, making the estimator inconsistent. The variance, however, decreases with the time series length so that the mean squared error remains quite stable with varying length of the series. The adjusted version of HXA outperforms the original absolute value based version in bias but variance-wise, it is the other way around. In terms of the mean squared error, the adjusted version outperforms the original one for this level of short term memory.

\autoref{tab:HXA6} summarizes the results for estimating the bivariate Hurst exponent for the mixed-correlated ARFIMA processes. We again observe, similarly to the DCCA case, that the estimator is upward biased. The bias increases with the time series length making the estimator inconsistent. Expectedly, the bias decreases with an increasing correlation between error terms. However, the decrease is rather mild. Connection of the bias to the maximum scale used varies for different time series lengths. The variance of the estimator decreases with time series length, increases with the maximum scale and remains practically unchanged for different levels of correlation between error terms. HXA estimator quite easily outperforms the DCCA approach for this type of processes. Note that we again report only the estimates for the original procedure as the method without the absolute values practically collapses in the same manner as for DCCA.

\subsection{Detrending moving-average cross-correlation analysis}

The last estimator we analyze is DMCA for which we vary the highest moving average window -- $\kappa_{max}=21,51,101$. The results of the Monte Carlo simulations are summarized in \autoref{tab:DMCA1}-\ref{tab:DMCA6_abs}. For weakly long-range cross-correlated processes based on correlated ARFIMA processes (\autoref{tab:DMCA1}), the estimator is unbiased regardless the time series length or other parameters. The variance of the estimator increases with the length of the applied moving average and decreases with the time series length for both specifications of the DMCA method. The variance is in general lower for the absolute value based estimator. Again, we do not report the results for weakly correlated ($\rho=0.1$) error terms for the original method as it provides only few actual estimates. For strongly long-range cross-correlated series (\autoref{tab:DMCA2}), the estimator is downward biased. The bias and variance decrease with the time series length for both specifications of the estimator hinting consistency. However, the rate of convergence seems rather slow as the decrease of bias is rather weak among the analyzed time series lengths. The bias also decreases with $\kappa_{max}$. Reversely, the variance increases with $\kappa_{max}$ and the total effect measured by the mean squared error varies with the time series length.

For the processes based on ARFIMA and AR(1), we observe that the short memory bias is not severe for weak short memory (\autoref{tab:DMCA3}) while there is slight downward bias for the original version of the estimator. The variance of the estimator increases with the moving average window size as does the mean squared error. The bias, variance and mean squared error decreases with time series length for all cases. For the medium and strong short memory (\autoref{tab:DMCA4}-\ref{tab:DMCA5}), the bias increases considerably and it becomes much higher than for the DCCA and HXA methods. Quite expectedly, the bias decreases with the maximum moving average window size for both medium ($\theta=0.5$) and strong memory ($\theta=0.8$). This is due to a stronger effect of short memory on the lower scales and thus lower moving average window sizes. In the same manner, the variance of the estimator increases with $\kappa_{max}$. The total effect results in the decreasing mean squared error with the maximum moving average size. Importantly, the DMCA estimates do not show a decreasing tendency of the bias with the increasing time series length.

For the mixed-correlated ARFIMA processes (\autoref{tab:DMCA6_abs}), we observe that the bias and variance decrease with the time series length but increase with $\kappa_{max}$. The bias also decreases with increasing correlation between error terms while the variance increases slightly. DMCA strongly outperforms both DCCA and HXA in this aspect.


\section{Conclusion}

We present a wide Monte Carlo simulation study focusing on various versions of power-law cross-correlations. Three estimators are analyzed, namely the detrended cross-correlation analysis (DCCA), height cross-correlation analysis (HXA) and detrending moving-average cross-correlation analysis (DMCA). We study various specifications of underlying processes including long-term memory, short-term memory and power-law coherency. Summarizing and comparing the performance of the three estimators, we find the following.

Firstly, the performance of all methods is dependent on the selection of parameters $s$, $\tau$ and $\kappa$. Secondly, the DCCA and HXA methods are downward biased for standard power-law cross-correlation case while DMCA reports satisfying results. Thirdly, DCCA outperforms the other two methods in resistance to the short-term memory bias and DMCA is by far the worst in this aspect. And fourthly, DMCA strongly outperforms the other two methods in the power-law coherency case. Each method is thus better suited for different types of processes and this should be taken into consideration when selecting the proper method for any analysis of power-law cross-correlations.

\section*{Acknowledgements}
The research leading to these results has received funding from the European Union's Seventh Framework Programme (FP7/2007-2013) under grant agreement No. FP7-SSH-612955 (FinMaP). Support from the Czech Science Foundation under project No. 14-11402P is also gratefully acknowledged.

\bibliography{Bibliography}
\bibliographystyle{unsrt}

\section*{Appendix}

\begin{table}[c]
\caption[Finite sample properties of DCCA I]{\textbf{Finite sample properties of DCCA I.} DCCA (\textit{top}) and DCCA$_{abs}$ (\textit{bottom}) estimators for correlated ARFIMA processes with $d_1=d_2=0.1$ and varying $\rho_{\varepsilon\nu}$.\label{tab:DCCA1}}
\centering
\footnotesize
\begin{tabular}{c|c|ccc|ccc|ccc}
\toprule \toprule
&&&$\rho=0.1$&&&$\rho=0.5$&&&$\rho=0.9$&\\
\midrule
&&bias&SD&MSE&bias&SD&MSE&bias&SD&MSE\\
\midrule \midrule
&$s_{min}=10$&-----&-----&-----&-0.0481&0.1255&0.0181&-0.0271&0.0717&0.0059\\
$T=500$&$s_{min}=20$&-----&-----&-----&-0.0519&0.1798&0.0350&-0.0229&0.0997&0.0105\\
&$s_{min}=50$&-----&-----&-----&-0.0717&0.3781&0.1481&-0.0215&0.1953&0.0386\\
\midrule
&$s_{min}=10$&-----&-----&-----&-0.0434&0.1060&0.0131&-0.0233&0.0558&0.0037\\
$T=1000$&$s_{min}=50$&-----&-----&-----&-0.0560&0.2148&0.0493&-0.0215&0.1098&0.0125\\
&$s_{min}=100$&-----&-----&-----&-0.0795&0.3817&0.1520&-0.0120&0.1852&0.0347\\
\midrule
&$s_{min}=10$&-----&-----&-----&-0.0249&0.0816&0.0073&-0.0135&0.0474&0.0024\\
$T=5000$&$s_{min}=50$&-----&-----&-----&-0.0275&0.1098&0.0128&-0.0121&0.0642&0.0043\\
&$s_{min}=100$&-----&-----&-----&-0.0314&0.1342&0.0190&-0.0130&0.0791&0.0064\\
\midrule \midrule
&$s_{min}=10$&-0.0285&0.0523&0.0036&-0.0287&0.0634&0.0048&-0.0266&0.0692&0.0055\\
$T=500$&$s_{min}=20$&-0.0245&0.0736&0.0060&-0.0253&0.0890&0.0086&-0.0223&0.0962&0.0098\\
&$s_{min}=50$&-0.0200&0.1455&0.0216&-0.0241&0.1711&0.0299&-0.0206&0.1887&0.0360\\
\midrule
&$s_{min}=10$&-0.0212&0.0426&0.0023&-0.0249&0.0524&0.0034&-0.0227&0.0538&0.0034\\
$T=1000$&$s_{min}=50$&-0.0155&0.0838&0.0073&-0.0240&0.1018&0.0109&-0.0205&0.1061&0.0117\\
&$s_{min}=100$&-0.0129&0.1440&0.0209&-0.0294&0.1663&0.0285&-0.0190&0.1786&0.0323\\
\midrule
&$s_{min}=10$&-0.0113&0.0340&0.0013&-0.0109&0.0415&0.0018&-0.0129&0.0457&0.0023\\
$T=5000$&$s_{min}=50$&-0.0098&0.0462&0.0022&-0.0090&0.0559&0.0032&-0.0114&0.0618&0.0040\\
&$s_{min}=100$&-0.0103&0.0572&0.0034&-0.0092&0.0684&0.0048&-0.0121&0.0762&0.0060\\
\bottomrule \bottomrule
\end{tabular}
\end{table}

\begin{table}[c]
\caption[Finite sample properties of DCCA II]{\textbf{Finite sample properties of DCCA II.} DCCA (\textit{top}) and DCCA$_{abs}$ (\textit{bottom})  estimators for correlated ARFIMA processes with $d_1=d_2=0.4$ and varying $\rho_{\varepsilon\nu}$.\label{tab:DCCA2}}
\centering
\footnotesize
\begin{tabular}{c|c|ccc|ccc|ccc}
\toprule \toprule
&&&$\rho=0.1$&&&$\rho=0.5$&&&$\rho=0.9$&\\
\midrule
&&bias&SD&MSE&bias&SD&MSE&bias&SD&MSE\\
\midrule \midrule
&$s_{min}=10$&-----&-----&-----&-0.0673&0.1655&0.0319&-0.0458&0.0861&0.0095\\
$T=500$&$s_{min}=20$&-----&-----&-----&-0.0684&0.2407&0.0626&-0.0394&0.1228&0.0166\\
&$s_{min}=50$&-----&-----&-----&-0.0938&0.6945&0.4911&-0.0399&0.2530&0.0656\\
\midrule
&$s_{min}=10$&-----&-----&-----&-0.0603&0.1314&0.0209&-0.0322&0.0750&0.0067\\
$T=1000$&$s_{min}=50$&-----&-----&-----&-0.0670&0.2551&0.0696&-0.0331&0.1448&0.0220\\
&$s_{min}=100$&-----&-----&-----&-0.0623&0.4987&0.2526&-0.0357&0.2582&0.0680\\
\midrule
&$s_{min}=10$&-----&-----&-----&-0.0433&0.1055&0.0130&-0.0221&0.0590&0.0040\\
$T=5000$&$s_{min}=50$&-----&-----&-----&-0.0511&0.1447&0.0235&-0.0216&0.0789&0.0067\\
&$s_{min}=100$&-----&-----&-----&-0.0610&0.1796&0.0360&-0.0243&0.0963&0.0099\\
\midrule \midrule
&$s_{min}=10$&-0.0452&0.0666&0.0065&-0.0411&0.0801&0.0081&-0.0439&0.0829&0.0088\\
$T=500$&$s_{min}=20$&-0.0394&0.0935&0.0103&-0.0340&0.1133&0.0140&-0.0376&0.1182&0.0154\\
&$s_{min}=50$&-0.0489&0.1961&0.0408&-0.0321&0.2360&0.0567&-0.0371&0.2430&0.0604\\
\midrule
&$s_{min}=10$&-0.0284&0.0515&0.0035&-0.0317&0.0648&0.0052&-0.0311&0.0717&0.0061\\
$T=1000$&$s_{min}=50$&-0.0246&0.1041&0.0114&-0.0234&0.1269&0.0167&-0.0310&0.1384&0.0201\\
&$s_{min}=100$&-0.0358&0.1883&0.0367&-0.0229&0.2148&0.0467&-0.0329&0.2475&0.0623\\
\midrule
&$s_{min}=10$&-0.0141&0.0435&0.0021&-0.0200&0.0510&0.0030&-0.0211&0.0566&0.0036\\
$T=5000$&$s_{min}=50$&-0.0120&0.0588&0.0036&-0.0200&0.0691&0.0052&-0.0203&0.0757&0.0061\\
&$s_{min}=100$&-0.0134&0.0727&0.0055&-0.0225&0.0848&0.0077&-0.0227&0.0924&0.0091\\
\bottomrule \bottomrule
\end{tabular}
\end{table}

\begin{table}[c]
\caption[Finite sample properties of DCCA III]{\textbf{Finite sample properties of DCCA III.} DCCA (\textit{top}) and DCCA$_{abs}$ (\textit{bottom})  estimators for correlated ARFIMA and AR(1) processes with $d=0.4$, $\theta=0.1$ and varying $\rho_{\varepsilon\nu}$.\label{tab:DCCA3}}
\centering
\footnotesize
\begin{tabular}{c|c|ccc|ccc|ccc}
\toprule \toprule
&&&$\rho=0.1$&&&$\rho=0.5$&&&$\rho=0.9$&\\
\midrule
&&bias&SD&MSE&bias&SD&MSE&bias&SD&MSE\\
\midrule \midrule
&$s_{min}=10$&-----&-----&-----&-0.0863&0.1669&0.0353&-0.0538&0.0829&0.0098\\
$T=500$&$s_{min}=20$&-----&-----&-----&-0.0944&0.2411&0.0671&-0.0503&0.1172&0.0163\\
&$s_{min}=50$&-----&-----&-----&-0.1190&0.5874&0.3592&-0.0450&0.2482&0.0636\\
\midrule
&$s_{min}=10$&-----&-----&-----&-0.0720&0.1310&0.0224&-0.0435&0.0692&0.0067\\
$T=1000$&$s_{min}=50$&-----&-----&-----&-0.0874&0.2683&0.0796&-0.0384&0.1400&0.0211\\
&$s_{min}=100$&-----&-----&-----&-0.1016&0.4775&0.2383&-0.0370&0.2413&0.0596\\
\midrule
&$s_{min}=10$&-----&-----&-----&-0.0502&0.1061&0.0138&-0.0277&0.0554&0.0038\\
$T=5000$&$s_{min}=50$&-----&-----&-----&-0.0548&0.1454&0.0241&-0.0263&0.0747&0.0063\\
&$s_{min}=100$&-----&-----&-----&-0.0607&0.1813&0.0366&-0.0283&0.0926&0.0094\\
\midrule \midrule
&$s_{min}=10$&-0.0172&0.0564&0.0035&-0.0271&0.0656&0.0050&-0.0332&0.0737&0.0065\\
$T=500$&$s_{min}=20$&-0.0167&0.0794&0.0066&-0.0255&0.0902&0.0088&-0.0297&0.1025&0.0114\\
&$s_{min}=50$&-0.0182&0.1573&0.0251&-0.0292&0.1780&0.0325&-0.0256&0.2044&0.0424\\
\midrule
&$s_{min}=10$&-0.0134&0.0450&0.0022&-0.0191&0.0523&0.0031&-0.0266&0.0603&0.0043\\
$T=1000$&$s_{min}=50$&-0.0116&0.0874&0.0078&-0.0198&0.1009&0.0106&-0.0220&0.1191&0.0147\\
&$s_{min}=100$&-0.0103&0.1533&0.0236&-0.0222&0.1734&0.0306&-0.0200&0.1988&0.0399\\
\midrule
&$s_{min}=10$&-0.0093&0.0363&0.0014&-0.0134&0.0424&0.0020&-0.0163&0.0472&0.0025\\
$T=5000$&$s_{min}=50$&-0.0101&0.0495&0.0026&-0.0136&0.0569&0.0034&-0.0154&0.0635&0.0043\\
&$s_{min}=100$&-0.0115&0.0618&0.0040&-0.0145&0.0696&0.0051&-0.0169&0.0784&0.0064\\
\bottomrule \bottomrule
\end{tabular}
\end{table}

\begin{table}[c]
\caption[Finite sample properties of DCCA IV]{\textbf{Finite sample properties of DCCA IV.} DCCA (\textit{top}) and DCCA$_{abs}$ (\textit{bottom}) estimators for correlated ARFIMA and AR(1) processes with $d=0.4$, $\theta=0.5$ and varying $\rho_{\varepsilon\nu}$.\label{tab:DCCA4}}
\centering
\footnotesize
\begin{tabular}{c|c|ccc|ccc|ccc}
\toprule \toprule
&&&$\rho=0.1$&&&$\rho=0.5$&&&$\rho=0.9$&\\
\midrule
&&bias&SD&MSE&bias&SD&MSE&bias&SD&MSE\\
\midrule \midrule
&$s_{min}=10$&-----&-----&-----&-0.0279&0.1599&0.0264&0.0006&0.0833&0.0069\\
$T=500$&$s_{min}=20$&-----&-----&-----&-0.0589&0.2318&0.0573&-0.0211&0.1156&0.0138\\
&$s_{min}=50$&-----&-----&-----&-0.1009&0.5401&0.3019&-0.0413&0.2420&0.0603\\
\midrule
&$s_{min}=10$&-----&-----&-----&-0.0328&0.1249&0.0167&-0.0137&0.0689&0.0049\\
$T=1000$&$s_{min}=50$&-----&-----&-----&-0.0703&0.2522&0.0686&-0.0462&0.1357&0.0205\\
&$s_{min}=100$&-----&-----&-----&-0.0791&0.4821&0.2386&-0.0574&0.2317&0.0570\\
\midrule
&$s_{min}=10$&-----&-----&-----&-0.0400&0.1048&0.0126&-0.0211&0.0516&0.0031\\
$T=5000$&$s_{min}=50$&-----&-----&-----&-0.0550&0.1436&0.0236&-0.0300&0.0701&0.0058\\
&$s_{min}=100$&-----&-----&-----&-0.0634&0.1781&0.0357&-0.0328&0.0870&0.0086\\
\midrule \midrule
&$s_{min}=10$&0.0442&0.0591&0.0054&0.0367&0.0698&0.0062&0.0214&0.0750&0.0061\\
$T=500$&$s_{min}=20$&0.0259&0.0834&0.0076&0.0190&0.0978&0.0099&0.0037&0.1030&0.0106\\
&$s_{min}=50$&0.0062&0.1712&0.02934&-0.0021&0.1914&0.0366&-0.0110&0.2071&0.0430\\
\midrule
&$s_{min}=10$&0.0337&0.0470&0.0033&0.0263&0.0545&0.0037&0.0074&0.0618&0.0039\\
$T=1000$&$s_{min}=50$&0.0085&0.0919&0.0085&0.0040&0.1059&0.0112&-0.0199&0.1189&0.0145\\
&$s_{min}=100$&0.0017&0.1569&0.0246&-0.0030&0.1751&0.0307&-0.0283&0.1935&0.0383\\
\midrule
&$s_{min}=10$&0.0132&0.0365&0.0015&0.0063&0.0419&0.0018&-0.0052&0.0454&0.0021\\
$T=5000$&$s_{min}=50$&0.0044&0.0499&0.0025&-0.0029&0.0565&0.0032&-0.0141&0.0613&0.0040\\
&$s_{min}=100$&0.0020&0.0624&0.0039&-0.0058&0.0692&0.0048&-0.0173&0.0755&0.0060\\
\bottomrule \bottomrule
\end{tabular}
\end{table}

\begin{table}[c]
\caption[Finite sample properties of DCCA V]{\textbf{Finite sample properties of DCCA V.} DCCA (\textit{top}) and DCCA$_{abs}$ (\textit{bottom}) estimators for correlated ARFIMA and AR(1) processes with $d=0.4$, $\theta=0.8$ and varying $\rho_{\varepsilon\nu}$.\label{tab:DCCA5}}
\centering
\footnotesize
\begin{tabular}{c|c|ccc|ccc|ccc}
\toprule \toprule
&&&$\rho=0.1$&&&$\rho=0.5$&&&$\rho=0.9$&\\
\midrule
&&bias&SD&MSE&bias&SD&MSE&bias&SD&MSE\\
\midrule \midrule
&$s_{min}=10$&-----&-----&-----&0.1326&0.1694&0.0463&0.1553&0.0877&0.0318\\
$T=500$&$s_{min}=20$&-----&-----&-----&0.0737&0.2427&0.0643&0.1021&0.1226&0.0255\\
&$s_{min}=50$&-----&-----&-----&-0.0219&0.5988&0.3590&0.0207&0.2548&0.0654\\
\midrule
&$s_{min}=10$&-----&-----&-----&0.0891&0.1226&0.0230&0.1027&0.0685&0.0152\\
$T=1000$&$s_{min}=50$&-----&-----&-----&-0.0061&0.2464&0.0607&0.0113&0.1356&0.0185\\
&$s_{min}=100$&-----&-----&-----&-0.0470&0.4676&0.2208&-0.0203&0.2327&0.0545\\
\midrule
&$s_{min}=10$&-----&-----&-----&-0.0033&0.1019&0.0104&0.0177&0.0544&0.0033\\
$T=5000$&$s_{min}=50$&-----&-----&-----&-0.0470&0.1408&0.0220&-0.0179&0.0730&0.0056\\
&$s_{min}=100$&-----&-----&-----&-0.0635&0.1761&0.0351&-0.0275&0.0890&0.0087\\
\midrule \midrule
&$s_{min}=10$&0.1687&0.0608&0.0321&0.1697&0.0758&0.0345&0.1603&0.0812&0.0323\\
$T=500$&$s_{min}=20$&0.1302&0.0857&0.0243&0.1310&0.1056&0.0283&0.1135&0.1134&0.0257\\
&$s_{min}=50$&0.0738&0.1810&0.0382&0.0716&0.2104&0.0494&0.0429&0.2295&0.0545\\
\midrule
&$s_{min}=10$&0.1303&0.0507&0.0196&0.1250&0.0593&0.0191&0.1133&0.0632&0.0168\\
$T=1000$&$s_{min}=50$&0.0620&0.1029&0.0144&0.0575&0.1125&0.0160&0.0337&0.1238&0.0165\\
&$s_{min}=100$&0.0306&0.1748&0.0315&0.0280&0.1922&0.0377&0.0079&0.2059&0.0424\\
\midrule
&$s_{min}=10$&0.0559&0.0374&0.0045&0.0467&0.0436&0.0041&0.0349&0.0485&0.0036\\
$T=5000$&$s_{min}=50$&0.0273&0.0503&0.0033&0.0162&0.0592&0.0038&0.0029&0.0649&0.0042\\
&$s_{min}=100$&0.0182&0.0617&0.0041&0.0064&0.0728&0.0053&-0.0059&0.0787&0.0062\\
\bottomrule \bottomrule
\end{tabular}
\end{table}

\begin{table}[c]
\caption[Finite sample properties of DCCA VI]{\textbf{Finite sample properties of DCCA VI.} $DCCA_{abs}$ estimator for Mixed-correlated ARFIMA processes with $d_1=d_4=0.4$, $d_2=d_3=0.2$ and varying $\rho_{\varepsilon\nu}$.\label{tab:DCCA6_abs}}
\centering
\footnotesize
\begin{tabular}{c|c|ccc|ccc|ccc}
\toprule \toprule
&&&$\rho=0.1$&&&$\rho=0.5$&&&$\rho=0.9$&\\
\midrule
&&bias&SD&MSE&bias&SD&MSE&bias&SD&MSE\\
\midrule \midrule
&$s_{min}=10$&0.0933&0.0628&0.0127&0.0853&0.0634&0.0113&0.0809&0.0652&0.0108\\
$T=500$&$s_{min}=20$&0.1076&0.0908&0.0198&0.0976&0.0897&0.0176&0.0943&0.0916&0.0173\\
&$s_{min}=50$&0.1248&0.1943&0.0533&0.1064&0.1872&0.0464&0.1035&0.1946&0.0486\\
\midrule
&$s_{min}=10$&0.1119&0.0515&0.01516&0.1066&0.0532&0.0142&0.1006&0.0563&0.0133\\
$T=1000$&$s_{min}=50$&0.1299&0.1025&0.0274&0.1247&0.1066&0.0269&0.1234&0.1094&0.0272\\
&$s_{min}=100$&0.1306&0.1823&0.0503&0.1187&0.1914&0.0507&0.1289&0.1094&0.0525\\
\midrule
&$s_{min}=10$&0.1409&0.0415&0.0216&0.1389&0.0407&0.0210&0.1348&0.0420&0.0199\\
$T=5000$&$s_{min}=50$&0.1494&0.0560&0.0255&0.1477&0.0551&0.0249&0.1449&0.0571&0.0243\\
&$s_{min}=100$&0.1511&0.0690&0.0276&0.1494&0.0681&0.0270&0.1475&0.0705&0.0267\\
\bottomrule \bottomrule
\end{tabular}
\end{table}

\begin{table}[c]
\caption[Finite sample properties of HXA I]{\textbf{Finite sample properties of HXA I.} HXA (\textit{top}) and HXA$_{abs}$ (\textit{bottom}) estimators for correlated ARFIMA processes with $d_1=d_2=0.1$ and varying $\rho_{\varepsilon\nu}$.\label{tab:HXA1}}
\centering
\footnotesize
\begin{tabular}{c|c|ccc|ccc|ccc}
\toprule \toprule
&&&$\rho=0.1$&&&$\rho=0.5$&&&$\rho=0.9$&\\
\midrule
&&bias&SD&MSE&bias&SD&MSE&bias&SD&MSE\\
\midrule \midrule
&$\tau_{max}=20$&-----&-----&-----&-0.0360&0.0721&0.0065&-0.0271&0.0423&0.0025\\
$T=500$&$\tau_{max}=50$&-----&-----&-----&-0.0524&0.0929&0.0114&-0.0357&0.0535&0.0041\\
&$\tau_{max}=100$&-----&-----&-----&-0.0732&0.1063&0.0167&-0.0557&0.0713&0.0082\\
\midrule
&$\tau_{max}=20$&-----&-----&-----&-0.0223&0.0446&0.0025&-0.0201&0.0286&0.0012\\
$T=1000$&$\tau_{max}=50$&-----&-----&-----&-0.0314&0.0608&0.0047&-0.0229&0.0352&0.0018\\
&$\tau_{max}=100$&-----&-----&-----&-0.0473&0.0802&0.0087&-0.0307&0.0451&0.0030\\
\midrule
&$\tau_{max}=20$&-----&-----&-----&-0.0147&0.0185&0.0006&-0.0135&0.0124&0.0003\\
$T=5000$&$\tau_{max}=50$&-----&-----&-----&-0.0134&0.0235&0.0007&-0.0110&0.0154&0.0004\\
&$\tau_{max}=100$&-----&-----&-----&-0.0150&0.0305&0.0012&-0.0104&0.0197&0.0005\\
\midrule \midrule
&$\tau_{max}=20$&-0.0234&0.0321&0.0016&-0.0251&0.0377&0.0020&-0.0267&0.0408&0.0024\\
$T=500$&$\tau_{max}=50$&-0.0304&0.0380&0.0024&-0.0318&0.0452&0.0031&-0.0348&0.0511&0.0038\\
&$\tau_{max}=100$&-0.0463&0.0465&0.0043&-0.0481&0.0561&0.0055&-0.0532&0.0665&0.0073\\
\midrule
&$\tau_{max}=20$&-0.0185&0.0218&0.0008&-0.0180&0.0264&0.0010&-0.0200&0.0275&0.0012\\
$T=1000$&$\tau_{max}=50$&-0.0193&0.0271&0.0011&-0.0208&0.0324&0.0015&-0.0225&0.0341&0.0017\\
&$\tau_{max}=100$&-0.0248&0.0341&0.0018&-0.0281&0.0406&0.0024&-0.0299&0.0433&0.0028\\
\midrule
&$\tau_{max}=20$&-0.0132&0.0105&0.0003&-0.0137&0.0111&0.0003&-0.0135&0.0121&0.0003\\
$T=5000$&$\tau_{max}=50$&-0.0110&0.0124&0.0003&-0.0117&0.0137&0.0003&-0.0110&0.0149&0.0003\\
&$\tau_{max}=100$&-0.0107&0.0151&0.0003&-0.0119&0.0174&0.0004&-0.0103&0.0190&0.0005\\
\bottomrule \bottomrule
\end{tabular}
\end{table}

\begin{table}[c]
\caption[Finite sample properties of HXA II]{\textbf{Finite sample properties of HXA II.} HXA (\textit{top}) and HXA$_{abs}$ (\textit{bottom}) estimators for correlated ARFIMA processes with $d_1=d_2=0.4$ and varying $\rho_{\varepsilon\nu}$.\label{tab:HXA2}}
\centering
\footnotesize
\begin{tabular}{c|c|ccc|ccc|ccc}
\toprule \toprule
&&&$\rho=0.1$&&&$\rho=0.5$&&&$\rho=0.9$&\\
\midrule
&&bias&SD&MSE&bias&SD&MSE&bias&SD&MSE\\
\midrule \midrule
&$\tau_{max}=20$&-----&-----&-----&-0.1103&0.0973&0.0216&-0.0865&0.0413&0.0092\\
$T=500$&$\tau_{max}=50$&-----&-----&-----&-0.1370&0.1234&0.0340&-0.1017&0.0525&0.0131\\
&$\tau_{max}=100$&-----&-----&-----&-0.1665&0.1333&0.0455&-0.1294&0.0704&0.0217\\
\midrule
&$\tau_{max}=20$&-----&-----&-----&-0.0816&0.0591&0.0101&-0.0700&0.0303&0.0058\\
$T=1000$&$\tau_{max}=50$&-----&-----&-----&-0.0971&0.0826&0.0162&-0.0774&0.0369&0.0074\\
&$\tau_{max}=100$&-----&-----&-----&-0.1191&0.1034&0.0249&-0.0911&0.0460&0.0104\\
\midrule
&$\tau_{max}=20$&-----&-----&-----&-0.0518&0.0281&0.0035&-0.0479&0.0180&0.0026\\
$T=5000$&$\tau_{max}=50$&-----&-----&-----&-0.0546&0.0344&0.0042&-0.0488&0.0212&0.0028\\
&$\tau_{max}=100$&-----&-----&-----&-0.0612&0.0442&0.0057&-0.0527&0.0251&0.0034\\
\midrule \midrule
&$\tau_{max}=20$&-0.0833&0.0335&0.0081&-0.0868&0.0356&0.0088&-0.0858&0.0395&0.0089\\
$T=500$&$\tau_{max}=50$&-0.0957&0.0397&0.0107&-0.1003&0.0435&0.0119&-0.1004&0.0496&0.0125\\
&$\tau_{max}=100$&-0.1186&0.0489&0.0164&-0.1240&0.0550&0.0184&-0.1267&0.0649&0.0203\\
\midrule
&$\tau_{max}=20$&-0.0678&0.0260&0.0053&-0.0700&0.0281&0.0057&-0.0697&0.0293&0.0057\\
$T=1000$&$\tau_{max}=50$&-0.0742&0.0307&0.0064&-0.0770&0.0334&0.0070&-0.0769&0.0354&0.0072\\
&$\tau_{max}=100$&-0.0865&0.0372&0.0887&-0.0895&0.0409&0.0097&-0.0903&0.0438&0.0101\\
\midrule
&$\tau_{max}=20$&-0.0473&0.0138&0.0024&-0.0477&0.0160&0.0025&-0.0478&0.0175&0.0026\\
$T=5000$&$\tau_{max}=50$&-0.0480&0.0163&0.0026&-0.0485&0.0185&0.0027&-0.0486&0.0205&0.0028\\
&$\tau_{max}=100$&-0.0514&0.0193&0.0030&-0.0522&0.0215&0.0032&-0.0524&0.0242&0.0033\\
\bottomrule \bottomrule
\end{tabular}
\end{table}

\begin{table}[c]
\caption[Finite sample properties of HXA III]{\textbf{Finite sample properties of HXA III.} HXA (\textit{top}) and HXA$_{abs}$ (\textit{bottom}) estimators for correlated ARFIMA and AR(1) processes with $d=0.4$, $\theta=0.1$ and varying $\rho_{\varepsilon\nu}$.\label{tab:HXA3}}
\centering
\footnotesize
\begin{tabular}{c|c|ccc|ccc|ccc}
\toprule \toprule
&&&$\rho=0.1$&&&$\rho=0.5$&&&$\rho=0.9$&\\
\midrule
&&bias&SD&MSE&bias&SD&MSE&bias&SD&MSE\\
\midrule \midrule
&$\tau_{max}=20$&-----&-----&-----&-0.0635&0.0856&0.0114&-0.0535&0.0440&0.0048\\
$T=500$&$\tau_{max}=50$&-----&-----&-----&-0.0854&0.1112&0.0197&-0.0665&0.0588&0.0079\\
&$\tau_{max}=100$&-----&-----&-----&-0.1044&0.1226&0.0259&-0.0929&0.0806&0.0151\\
\midrule
&$\tau_{max}=20$&-----&-----&-----&-0.0531&0.0602&0.0064&-0.0416&0.0309&0.0027\\
$T=1000$&$\tau_{max}=50$&-----&-----&-----&-0.0694&0.0876&0.0125&-0.0455&0.0396&0.0036\\
&$\tau_{max}=100$&-----&-----&-----&-0.0885&0.1065&0.00192&-0.0566&0.0531&0.0060\\
\midrule
&$\tau_{max}=20$&-----&-----&-----&-0.0330&0.0247&0.0017&-0.0308&0.0142&0.0012\\
$T=5000$&$\tau_{max}=50$&-----&-----&-----&-0.0324&0.0325&0.0021&-0.0280&0.0182&0.0011\\
&$\tau_{max}=100$&-----&-----&-----&-0.0356&0.0441&0.0032&-0.0275&0.0227&0.0013\\
\midrule \midrule
&$\tau_{max}=20$&-0.0278&0.0316&0.0018&-0.0309&0.0339&0.0021&-0.0392&0.0377&0.0030\\
$T=500$&$\tau_{max}=50$&-0.0426&0.0386&0.0033&-0.0452&0.0418&0.0038&-0.0531&0.0475&0.0051\\
&$\tau_{max}=100$&-0.0643&0.0487&0.0065&-0.0673&0.0544&0.0075&-0.0767&0.0623&0.0098\\
\midrule
&$\tau_{max}=20$&-0.0190&0.0225&0.0009&-0.0239&0.0254&0.0012&-0.0278&0.0267&0.0015\\
$T=1000$&$\tau_{max}=50$&-0.0288&0.0264&0.0015&-0.0337&0.0303&0.0021&-0.0350&0.0331&0.0023\\
&$\tau_{max}=100$&-0.0414&0.0326&0.0028&-0.0464&0.0376&0.0036&-0.0466&0.0417&0.0039\\
\midrule
&$\tau_{max}=20$&-0.0065&0.0115&0.0002&-0.0105&0.0122&0.0003&-0.0162&0.0131&0.0004\\
$T=5000$&$\tau_{max}=50$&-0.0120&0.0133&0.0003&-0.0153&0.0147&0.0005&-0.0184&0.0159&0.0006\\
&$\tau_{max}=100$&-0.0174&0.0159&0.0006&-0.0202&0.0182&0.0007&-0.0214&0.0194&0.0008\\
\bottomrule \bottomrule
\end{tabular}
\end{table}

\begin{table}[c]
\caption[Finite sample properties of HXA IV]{\textbf{Finite sample properties of HXA IV.} HXA (\textit{top}) and HXA$_{abs}$ (\textit{bottom}) estimators for correlated ARFIMA and AR(1) processes with $d=0.4$, $\theta=0.5$ and varying $\rho_{\varepsilon\nu}$.\label{tab:HXA4}}
\centering
\footnotesize
\begin{tabular}{c|c|ccc|ccc|ccc}
\toprule \toprule
&&&$\rho=0.1$&&&$\rho=0.5$&&&$\rho=0.9$&\\
\midrule
&&bias&SD&MSE&bias&SD&MSE&bias&SD&MSE\\
\midrule \midrule
&$\tau_{max}=20$&-----&-----&-----&0.0082&0.0837&0.0071&0.0226&0.0403&0.0021\\
$T=500$&$\tau_{max}=50$&-----&-----&-----&-0.0390&0.1099&0.0136&-0.0121&0.0568&0.0034\\
&$\tau_{max}=100$&-----&-----&-----&-0.0750&0.1191&0.0198&-0.0505&0.0784&0.0087\\
\midrule
&$\tau_{max}=20$&-----&-----&-----&0.0230&0.0518&0.0032&0.0312&0.0273&0.0017\\
$T=1000$&$\tau_{max}=50$&-----&-----&-----&-0.0162&0.0801&0.0067&0.0029&0.0374&0.0014\\
&$\tau_{max}=100$&-----&-----&-----&-0.0503&0.1013&0.0128&-0.0232&0.0514&0.0032\\
\midrule
&$\tau_{max}=20$&-----&-----&-----&0.0394&0.0216&0.0020&0.0409&0.0124&0.0018\\
$T=5000$&$\tau_{max}=50$&-----&-----&-----&0.0161&0.0300&0.0012&0.0195&0.0166&0.0007\\
&$\tau_{max}=100$&-----&-----&-----&-0.0015&0.0414&0.0017&0.0042&0.0216&0.0005\\
\midrule \midrule
&$\tau_{max}=20$&0.0515&0.0304&0.0036&0.0452&0.0333&0.0032&0.0368&0.0365&0.0023\\
$T=500$&$\tau_{max}=50$&0.0167&0.0366&0.0016&0.0111&0.0422&0.0019&0.0035&0.0481&0.0023\\
&$\tau_{max}=100$&-0.0188&0.0469&0.0026&-0.0253&0.0540&0.0036&-0.0322&0.0630&0.0050\\
\midrule
&$\tau_{max}=20$&0.0604&0.0222&0.0041&0.0538&0.0246&0.0035&0.0459&0.0250&0.0027\\
$T=1000$&$\tau_{max}=50$&0.0306&0.0276&0.0017&0.0247&0.0300&0.0015&0.0174&0.0323&0.0013\\
&$\tau_{max}=100$&0.0046&0.0339&0.0012&-0.0018&0.0367&0.0013&-0.0083&0.0419&0.0018\\
\midrule
&$\tau_{max}=20$&0.0717&0.0119&0.0053&0.0672&0.0121&0.0047&0.0577&0.0121&0.0035\\
$T=5000$&$\tau_{max}=50$&0.0464&0.0138&0.0023&0.0425&0.0146&0.0020&0.0346&0.0151&0.0014\\
&$\tau_{max}=100$&0.0269&0.0163&0.0010&0.0236&0.0177&0.0009&0.0171&0.0189&0.0006\\
\bottomrule \bottomrule
\end{tabular}
\end{table}

\begin{table}[c]
\caption[Finite sample properties of HXA V]{\textbf{Finite sample properties of HXA V.} HXA (\textit{top}) and HXA$_{abs}$ (\textit{bottom}) estimators for correlated ARFIMA and AR(1) processes with $d=0.4$, $\theta=0.8$ and varying $\rho_{\varepsilon\nu}$.\label{tab:HXA5}}
\centering
\footnotesize
\begin{tabular}{c|c|ccc|ccc|ccc}
\toprule \toprule
&&&$\rho=0.1$&&&$\rho=0.5$&&&$\rho=0.9$&\\
\midrule
&&bias&SD&MSE&bias&SD&MSE&bias&SD&MSE\\
\midrule \midrule
&$\tau_{max}=20$&-----&-----&-----&0.1137&0.0687&0.0176&0.1263&0.0320&0.0170\\
$T=500$&$\tau_{max}=50$&-----&-----&-----&0.0538&0.0959&0.0121&0.0780&0.0484&0.0084\\
&$\tau_{max}=100$&-----&-----&-----&-0.0019&0.1106&0.0122&0.0213&0.0710&0.0055\\
\midrule
&$\tau_{max}=20$&-----&-----&-----&0.1290&0.0417&0.0184&0.1350&0.0210&0.0187\\
$T=1000$&$\tau_{max}=50$&-----&-----&-----&0.0788&0.0681&0.0108&0.0928&0.0315&0.0096\\
&$\tau_{max}=100$&-----&-----&-----&0.0285&0.0916&0.0092&0.0498&0.0457&0.0046\\
\midrule
&$\tau_{max}=20$&-----&-----&-----&0.1424&0.0156&0.0205&0.1427&0.0098&0.0205\\
$T=5000$&$\tau_{max}=50$&-----&-----&-----&0.1052&0.0228&0.0116&0.1064&0.0143&0.0115\\
&$\tau_{max}=100$&-----&-----&-----&0.0707&0.0336&0.0061&0.0735&0.0201&0.0058\\
\midrule \midrule
&$\tau_{max}=20$&0.1321&0.0271&0.0182&0.1308&0.0295&0.0180&0.1290&0.0307&0.0176\\
$T=500$&$\tau_{max}=50$&0.0921&0.0358&0.0098&0.0881&0.0395&0.0093&0.0857&0.0429&0.0092\\
&$\tau_{max}=100$&0.0461&0.0479&0.0044&0.0396&0.0525&0.0043&0.0357&0.0584&0.0047\\
\midrule
&$\tau_{max}=20$&0.1400&0.0203&0.0200&0.1395&0.0211&0.0199&0.1381&0.0207&0.0195\\
$T=1000$&$\tau_{max}=50$&0.1041&0.0256&0.0115&0.1030&0.0274&0.0114&0.0995&0.0295&0.0108\\
&$\tau_{max}=100$&0.0675&0.0323&0.0056&0.0657&0.0351&0.0056&0.0602&0.0400&0.0052\\
\midrule
&$\tau_{max}=20$&0.1533&0.0105&0.0236&0.1513&0.0105&0.0230&0.1480&0.0104&0.0220\\
$T=5000$&$\tau_{max}=50$&0.1221&0.0129&0.0151&0.1196&0.0132&0.0145&0.1148&0.0140&0.0134\\
&$\tau_{max}=100$&0.0920&0.0159&0.0097&0.0896&0.0166&0.0083&0.0837&0.0185&0.0070\\
\bottomrule \bottomrule
\end{tabular}
\end{table}

\begin{table}[c]
\caption[Finite sample properties of HXA VI]{\textbf{Finite sample properties of HXA VI.} HXA$_{abs}$ estimator for Mixed-correlated ARFIMA processes with $d_1=d_4=0.4$, $d_2=d_3=0.2$ and varying $\rho_{\varepsilon\nu}$.\label{tab:HXA6}}
\centering
\footnotesize
\begin{tabular}{c|c|ccc|ccc|ccc}
\toprule \toprule
&&&$\rho=0.1$&&&$\rho=0.5$&&&$\rho=0.9$&\\
\midrule
&&bias&SD&MSE&bias&SD&MSE&bias&SD&MSE\\
\midrule \midrule
&$\tau_{max}=20$&0.0678&0.0365&0.0059&0.0624&0.0354&0.0052&0.0564&0.0363&0.0045\\
$T=500$&$\tau_{max}=50$&0.0615&0.0434&0.0057&0.0566&0.0430&0.0051&0.0504&0.0442&0.0045\\
&$\tau_{max}=100$&0.0447&0.0527&0.0048&0.0395&0.0537&0.0044&0.0331&0.0560&0.0042\\
\midrule
&$\tau_{max}=20$&0.0812&0.0275&0.0074&0.0798&0.0253&0.0070&0.0729&0.0271&0.0060\\
$T=1000$&$\tau_{max}=50$&0.0815&0.0320&0.0077&0.0802&0.0302&0.0073&0.0743&0.0320&0.0065\\
&$\tau_{max}=100$&0.0744&0.0388&0.0070&0.0740&0.0369&0.0068&0.0685&0.0387&0.0062\\
\midrule
&$\tau_{max}=20$&0.1033&0.0143&0.0109&0.1012&0.0151&0.0105&0.0946&0.0150&0.0092\\
$T=5000$&$\tau_{max}=50$&0.1091&0.0167&0.0122&0.1075&0.0173&0.0119&0.1015&0.0176&0.0106\\
&$\tau_{max}=100$&0.1111&0.0197&0.0127&0.1098&0.0201&0.0125&0.1043&0.0207&0.0113\\
\bottomrule \bottomrule
\end{tabular}
\end{table}

\begin{table}[c]
\caption[Finite sample properties of DMCA I]{\textbf{Finite sample properties of DMCA I.} DMCA (\textit{top}) and DMCA$_{abs}$ (\textit{bottom}) estimators for correlated ARFIMA processes with $d_1=d_2=0.1$ and varying $\rho_{\varepsilon\nu}$.\label{tab:DMCA1}}
\centering
\footnotesize
\begin{tabular}{c|c|ccc|ccc|ccc}
\toprule \toprule
&&&$\rho=0.1$&&&$\rho=0.5$&&&$\rho=0.9$&\\
\midrule
&&bias&SD&MSE&bias&SD&MSE&bias&SD&MSE\\
\midrule \midrule
&$\kappa_{max}=21$&-----&-----&-----&-0.0050&0.0641&0.0041&-0.0008&0.0410&0.0017\\
$T=500$&$\kappa_{max}=51$&-----&-----&-----&-0.0118&0.0787&0.0063&-0.0064&0.0484&0.0024\\
&$\kappa_{max}=101$&-----&-----&-----&-0.0237&0.1017&0.0109&-0.0113&0.0599&0.0037\\
\midrule
&$\kappa_{max}=21$&-----&-----&-----&-0.0066&0.0466&0.0022&-0.0030&0.0304&0.0009\\
$T=1000$&$\kappa_{max}=51$&-----&-----&-----&-0.0151&0.0537&0.0031&-0.0077&0.0350&0.0013\\
&$\kappa_{max}=101$&-----&-----&-----&-0.0194&0.0710&0.0054&-0.0074&0.0436&0.0020\\
\midrule
&$\kappa_{max}=21$&-----&-----&-----&-0.0021&0.0203&0.0004&-0.0021&0.0134&0.0002\\
$T=5000$&$\kappa_{max}=51$&-----&-----&-----&-0.0058&0.0228&0.0006&-0.0054&0.0157&0.0003\\
&$\kappa_{max}=101$&-----&-----&-----&-0.0066&0.0277&0.0008&-0.0058&0.0190&0.0004\\
\midrule \midrule
&$\kappa_{max}=21$&-0.0018&0.0331&0.0011&-0.0031&0.0381&0.0015&-0.0008&0.0397&0.0016\\
$T=500$&$\kappa_{max}=51$&-0.0059&0.0376&0.0015&-0.0067&0.0438&0.0020&-0.0063&0.0467&0.0022\\
&$\kappa_{max}=101$&-0.0090&0.0462&0.0022&-0.0105&0.0531&0.0029&-0.0109&0.0578&0.0035\\
\midrule
&$\kappa_{max}=21$&-0.0030&0.0217&0.0005&-0.0044&0.0273&0.0008&-0.0030&0.0294&0.0009\\
$T=1000$&$\kappa_{max}=51$&-0.0072&0.0252&0.0007&-0.0095&0.0300&0.0010&-0.0075&0.0339&0.0012\\
&$\kappa_{max}=101$&-0.0071&0.0319&0.0011&-0.0095&0.0377&0.0015&-0.0073&0.0422&0.0018\\
\midrule
&$\kappa_{max}=21$&-0.0025&0.0102&0.0001&-0.0017&0.0118&0.0001&-0.0021&0.0130&0.0002\\
$T=5000$&$\kappa_{max}=51$&-0.0057&0.0115&0.0002&-0.0053&0.0135&0.0002&-0.0054&0.0152&0.0003\\
&$\kappa_{max}=101$&-0.0054&0.0139&0.0002&-0.0057&0.0164&0.0003&-0.0058&0.0184&0.0004\\
\bottomrule \bottomrule
\end{tabular}
\end{table}

\begin{table}[c]
\caption[Finite sample properties of DMCA II]{\textbf{Finite sample properties of DMCA II.} DMCA (\textit{top}) and DMCA$_{abs}$ (\textit{bottom}) estimators for correlated ARFIMA processes with $d_1=d_2=0.4$ and varying $\rho_{\varepsilon\nu}$.\label{tab:DMCA2}}
\centering
\footnotesize
\begin{tabular}{c|c|ccc|ccc|ccc}
\toprule \toprule
&&&$\rho=0.1$&&&$\rho=0.5$&&&$\rho=0.9$&\\
\midrule
&&bias&SD&MSE&bias&SD&MSE&bias&SD&MSE\\
\midrule \midrule
&$\kappa_{max}=21$&-----&-----&-----&-0.0613&0.0718&0.0089&-0.0593&0.0471&0.0057\\
$T=500$&$\kappa_{max}=51$&-----&-----&-----&-0.0508&0.0992&0.0124&-0.0411&0.0583&0.0051\\
&$\kappa_{max}=101$&-----&-----&-----&-0.0557&0.1358&0.02156&-0.0344&0.0728&0.0065\\
\midrule
&$\kappa_{max}=21$&-----&-----&-----&-0.0583&0.0502&0.0059&-0.0550&0.0314&0.0040\\
$T=1000$&$\kappa_{max}=51$&-----&-----&-----&-0.0407&0.0613&0.0054&-0.0366&0.0400&0.0029\\
&$\kappa_{max}=101$&-----&-----&-----&-0.0367&0.0850&0.0086&-0.0266&0.0516&0.0034\\
\midrule
&$\kappa_{max}=21$&-----&-----&-----&-0.0569&0.0221&0.0037&-0.0563&0.0141&0.0034\\
$T=5000$&$\kappa_{max}=51$&-----&-----&-----&-0.0374&0.0258&0.0021&-0.0349&0.0168&0.0015\\
&$\kappa_{max}=101$&-----&-----&-----&-0.0255&0.0313&0.0016&-0.0233&0.0210&0.0010\\
\midrule \midrule
&$\kappa_{max}=21$&-0.0574&0.0371&0.0047&-0.0586&0.0432&0.0053&-0.0592&0.0457&0.0056\\
$T=500$&$\kappa_{max}=51$&-0.0395&0.0421&0.0033&-0.0406&0.0530&0.0045&-0.0407&0.0564&0.0048\\
&$\kappa_{max}=101$&-0.0310&0.0547&0.0040&-0.0320&0.0654&0.0053&-0.0335&0.0699&0.0060\\
\midrule
&$\kappa_{max}=21$&-0.0553&0.0241&0.0036&-0.0562&0.0289&0.0040&-0.0550&0.0304&0.0040\\
$T=1000$&$\kappa_{max}=51$&-0.0355&0.0292&0.0021&-0.0360&0.0346&0.0025&-0.0366&0.0388&0.0028\\
&$\kappa_{max}=101$&-0.0245&0.0372&0.0020&-0.0265&0.0452&0.0027&-0.0264&0.0499&0.0032\\
\midrule
&$\kappa_{max}=21$&-0.0564&0.0111&0.0033&-0.0564&0.0132&0.0034&-0.0562&0.0137&0.0033\\
$T=5000$&$\kappa_{max}=51$&-0.0355&0.0134&0.0014&-0.0358&0.0151&0.0015&-0.0348&0.0163&0.0015\\
&$\kappa_{max}=101$&-0.0240&0.0164&0.0008&-0.0239&0.0188&0.0009&-0.0233&0.0204&0.0010\\
\bottomrule \bottomrule
\end{tabular}
\end{table}

\begin{table}[c]
\caption[Finite sample properties of DMCA III]{\textbf{Finite sample properties of DMCA III.} DMCA (\textit{top}) and DMCA$_{abs}$ (\textit{bottom}) estimators for correlated ARFIMA and AR(1) processes with $d=0.4$, $\theta=0.1$ and varying $\rho_{\varepsilon\nu}$.\label{tab:DMCA3}}
\centering
\footnotesize
\begin{tabular}{c|c|ccc|ccc|ccc}
\toprule \toprule
&&&$\rho=0.1$&&&$\rho=0.5$&&&$\rho=0.9$&\\
\midrule
&&bias&SD&MSE&bias&SD&MSE&bias&SD&MSE\\
\midrule \midrule
&$\kappa_{max}=21$&-----&-----&-----&-0.0274&0.0741&0.0062&-0.0278&0.0453&0.0028\\
$T=500$&$\kappa_{max}=51$&-----&-----&-----&-0.0364&0.0908&0.0106&-0.0316&0.0514&0.0036\\
&$\kappa_{max}=101$&-----&-----&-----&-0.0524&0.1349&0.0209&-0.0330&0.0680&0.0057\\
\midrule
&$\kappa_{max}=21$&-----&-----&-----&-0.0260&0.0487&0.0030&-0.0247&0.0325&0.0017\\
$T=1000$&$\kappa_{max}=51$&-----&-----&-----&-0.0321&0.0621&0.0049&-0.0269&0.0381&0.0022\\
&$\kappa_{max}=101$&-----&-----&-----&-0.0375&0.0875&0.0091&-0.0259&0.0462&0.0028\\
\midrule
&$\kappa_{max}=21$&-----&-----&-----&-0.0246&0.0225&0.0011&-0.0238&0.0138&0.0008\\
$T=5000$&$\kappa_{max}=51$&-----&-----&-----&-0.0274&0.0277&0.0015&-0.0263&0.0169&0.0010\\
&$\kappa_{max}=101$&-----&-----&-----&-0.0256&0.0352&0.0019&-0.0236&0.0207&0.0010\\
\midrule \midrule
&$\kappa_{max}=21$&0.0125&0.0339&0.0013&0.0065&0.0389&0.0016&-0.0106&0.0424&0.0019\\
$T=500$&$\kappa_{max}=51$&0.0063&0.0372&0.0014&0.0010&0.0448&0.0020&-0.0139&0.0472&0.0024\\
&$\kappa_{max}=101$&0.0013&0.0485&0.0024&-0.0040&0.0550&0.0030&-0.0159&0.0611&0.0040\\
\midrule
&$\kappa_{max}=21$&0.0152&0.0233&0.0008&0.0065&0.0259&0.0007&-0.0080&0.0305&0.0010\\
$T=1000$&$\kappa_{max}=51$&0.0076&0.0274&0.0008&0.0019&0.0305&0.0009&-0.0101&0.0350&0.0013\\
&$\kappa_{max}=101$&0.0030&0.0347&0.0012&-0.0008&0.0374&0.0014&-0.0105&0.0424&0.0019\\
\midrule
&$\kappa_{max}=21$&0.0148&0.0106&0.0003&0.0073&0.0122&0.0002&-0.0074&0.0131&0.0002\\
$T=5000$&$\kappa_{max}=51$&0.0089&0.0120&0.0002&0.0026&0.0140&0.0002&-0.0099&0.0156&0.0003\\
&$\kappa_{max}=101$&0.0059&0.0143&0.0002&0.0011&0.0168&0.0003&-0.0089&0.0188&0.0004\\
\bottomrule \bottomrule
\end{tabular}
\end{table}

\begin{table}[c]
\caption[Finite sample properties of DMCA IV]{\textbf{Finite sample properties of DMCA IV.} DMCA (\textit{top}) and DMCA$_{abs}$ (\textit{bottom}) estimators for correlated ARFIMA and AR(1) processes with $d=0.4$, $\theta=0.5$ and varying $\rho_{\varepsilon\nu}$.\label{tab:DMCA4}}
\centering
\footnotesize
\begin{tabular}{c|c|ccc|ccc|ccc}
\toprule \toprule
&&&$\rho=0.1$&&&$\rho=0.5$&&&$\rho=0.9$&\\
\midrule
&&bias&SD&MSE&bias&SD&MSE&bias&SD&MSE\\
\midrule \midrule
&$\kappa_{max}=21$&-----&-----&-----&0.1493&0.0725&0.0275&0.1503&0.0478&0.0249\\
$T=500$&$\kappa_{max}=51$&-----&-----&-----&0.0797&0.0967&0.0157&0.0891&0.0521&0.0107\\
&$\kappa_{max}=101$&-----&-----&-----&0.0187&0.1344&0.0184&0.0447&0.0699&0.0069\\
\midrule
&$\kappa_{max}=21$&-----&-----&-----&0.1481&0.0485&0.0243&0.1493&0.0324&0.0233\\
$T=1000$&$\kappa_{max}=51$&-----&-----&-----&0.0857&0.0657&0.0117&0.0903&0.0366&0.0095\\
&$\kappa_{max}=101$&-----&-----&-----&0.0433&0.0875&0.0095&0.0487&0.0450&0.0044\\
\midrule
&$\kappa_{max}=21$&-----&-----&-----&0.1502&0.0221&0.0230&0.1521&0.0144&0.0233\\
$T=5000$&$\kappa_{max}=51$&-----&-----&-----&0.0927&0.0269&0.0093&0.0933&0.0167&0.0090\\
&$\kappa_{max}=101$&-----&-----&-----&0.0548&0.0322&0.0040&0.0546&0.0199&0.0034\\
\midrule \midrule
&$\kappa_{max}=21$&0.1632&0.0347&0.0278&0.1623&0.0417&0.0281&0.1549&0.0458&0.0261\\
$T=500$&$\kappa_{max}=51$&0.1172&0.0383&0.0152&0.1114&0.0480&0.0147&0.1003&0.0492&0.0125\\
&$\kappa_{max}=101$&0.0832&0.0487&0.0093&0.0726&0.0559&0.0084&0.0604&0.0651&0.0079\\
\midrule
&$\kappa_{max}=21$&0.1636&0.0253&0.0274&0.1603&0.0283&0.0265&0.1538&0.0312&0.0246\\
$T=1000$&$\kappa_{max}=51$&0.1186&0.0287&0.0149&0.1119&0.0337&0.0137&0.1009&0.0348&0.0114\\
&$\kappa_{max}=101$&0.0843&0.0337&0.0082&0.0790&0.0405&0.0079&0.0632&0.0419&0.0058\\
\midrule
&$\kappa_{max}=21$&0.1648&0.0112&0.0273&0.1603&0.0127&0.0259&0.1564&0.0138&0.0247\\
$T=5000$&$\kappa_{max}=51$&0.1198&0.0124&0.0145&0.1139&0.0150&0.0132&0.1034&0.0158&0.0109\\
&$\kappa_{max}=101$&0.0867&0.0149&0.0077&0.0817&0.0169&0.0070&0.0680&0.0187&0.0050\\
\bottomrule \bottomrule
\end{tabular}
\end{table}

\begin{table}[c]
\caption[Finite sample properties of DMCA V]{\textbf{Finite sample properties of DMCA V.} DMCA (\textit{top}) and DMCA$_{abs}$ (\textit{bottom}) estimators for correlated ARFIMA and AR(1) processes with $d=0.4$, $\theta=0.8$ and varying $\rho_{\varepsilon\nu}$.\label{tab:DMCA5}}
\centering
\footnotesize
\begin{tabular}{c|c|ccc|ccc|ccc}
\toprule \toprule
&&&$\rho=0.1$&&&$\rho=0.5$&&&$\rho=0.9$&\\
\midrule
&&bias&SD&MSE&bias&SD&MSE&bias&SD&MSE\\
\midrule \midrule
&$\kappa_{max}=21$&-----&-----&-----&0.3160&0.0759&0.1056&0.3225&0.0472&0.1062\\
$T=500$&$\kappa_{max}=51$&-----&-----&-----&0.2717&0.0939&0.0826&0.2761&0.0557&0.0793\\
&$\kappa_{max}=101$&-----&-----&-----&0.1920&0.1324&0.0544&0.2087&0.0700&0.0485\\
\midrule
&$\kappa_{max}=21$&-----&-----&-----&0.3288&0.0534&0.1078&0.3255&0.0339&0.1071\\
$T=1000$&$\kappa_{max}=51$&-----&-----&-----&0.2771&0.0618&0.0806&0.2837&0.0388&0.0820\\
&$\kappa_{max}=101$&-----&-----&-----&0.2074&0.0799&0.0494&0.2171&0.0481&0.0493\\
\midrule
&$\kappa_{max}=21$&-----&-----&-----&0.3268&0.0229&0.1074&0.3269&0.0151&0.1071\\
$T=5000$&$\kappa_{max}=51$&-----&-----&-----&0.2843&0.0264&0.0815&0.2853&0.0172&0.0817\\
&$\kappa_{max}=101$&-----&-----&-----&0.2181&0.0329&0.0487&0.2204&0.0204&0.0490\\
\midrule \midrule
&$\kappa_{max}=21$&0.3130&0.0362&0.0999&0.3131&0.0426&0.0999&0.3184&0.0447&0.1034\\
$T=500$&$\kappa_{max}=51$&0.2667&0.0409&0.0728&0.2700&0.0507&0.0755&0.2718&0.0530&0.0767\\
&$\kappa_{max}=101$&0.2101&0.0526&0.0469&0.2116&0.0628&0.0487&0.2099&0.0661&0.0484\\
\midrule
&$\kappa_{max}=21$&0.3155&0.0236&0.1002&0.3177&0.0304&0.1018&0.3212&0.0321&0.1042\\
$T=1000$&$\kappa_{max}=51$&0.2728&0.0303&0.0753&0.2730&0.0344&0.0757&0.2789&0.0372&0.0792\\
&$\kappa_{max}=101$&0.2189&0.0366&0.0493&0.2164&0.0426&0.0486&0.2174&0.0448&0.0493\\
\midrule
&$\kappa_{max}=21$&0.3157&0.0114&0.0998&0.3187&0.0130&0.1017&0.3225&0.0143&0.1042\\
$T=5000$&$\kappa_{max}=51$&0.2719&0.0133&0.0741&0.2746&0.0150&0.0756&0.2802&0.0165&0.0788\\
&$\kappa_{max}=101$&0.2197&0.0157&0.0485&0.2186&0.0180&0.0481&0.2200&0.0195&0.0488\\
\bottomrule \bottomrule
\end{tabular}
\end{table}

\begin{table}[c]
\caption[Finite sample properties of DMCA VI]{\textbf{Finite sample properties of DMCA VI.} DMCA$_{abs}$ estimator for Mixed-correlated ARFIMA processes with $d_1=d_4=0.4$, $d_2=d_3=0.2$ and varying $\rho_{\varepsilon\nu}$.\label{tab:DMCA6_abs}}
\centering
\footnotesize
\begin{tabular}{c|c|ccc|ccc|ccc}
\toprule \toprule
&&&$\rho=0.1$&&&$\rho=0.5$&&&$\rho=0.9$&\\
\midrule
&&bias&SD&MSE&bias&SD&MSE&bias&SD&MSE\\
\midrule \midrule
&$\kappa_{max}=21$&0.0649&0.0345&0.0054&0.0604&0.0365&0.0049&0.0499&0.0379&0.0039\\
$T=500$&$\kappa_{max}=51$&0.0864&0.0399&0.0091&0.0840&0.0419&0.0088&0.0715&0.0436&0.0070\\
&$\kappa_{max}=101$&0.1012&0.0536&0.0131&0.0983&0.0521&0.0124&0.0881&0.0544&0.0107\\
\midrule
&$\kappa_{max}=21$&0.0640&0.0241&0.0047&0.0596&0.0248&0.0042&0.0515&0.0282&0.0034\\
$T=1000$&$\kappa_{max}=51$&0.0882&0.0295&0.0086&0.0828&0.0288&0.0077&0.0751&0.0310&0.0066\\
&$\kappa_{max}=101$&0.1055&0.0362&0.0124&0.0998&0.0348&0.0112&0.0925&0.0382&0.0100\\
\midrule
&$\kappa_{max}=21$&0.0645&0.0106&0.0043&0.0610&0.0108&0.0038&0.0509&0.0123&0.0027\\
$T=5000$&$\kappa_{max}=51$&0.0892&0.0128&0.0081&0.0850&0.0123&0.0074&0.0756&0.0138&0.0059\\
&$\kappa_{max}=101$&0.1069&0.0154&0.0117&0.1022&0.0157&0.0107&0.0949&0.0174&0.0093\\
\bottomrule \bottomrule
\end{tabular}
\end{table}

\end{document}